\documentclass[12pt]{article}
\usepackage{a4}
\usepackage{epsfig}
\begin{document}
\title {Enhanced radiometric forces}
\author{Marco Scandurra \thanks{Present address: scandurr@lns.mit.edu}\\
{\itshape Center for Theoretical Physics}\\
{\itshape Massachusetts Institute of Technology}\\
{\itshape 77 Massachusetts Avenue, Cambridge, MA 02139}}\maketitle

\begin{abstract}

Crookes' Radiometer exploits thermal forces arising from temperature gradients in rarefied gases. This instrument has remained
confined to educational physics  because of the very small size of the radiometric effect and because of the necessity of an evacuated environment for operation. In this article the  physics of this instrument is reviewed, different theories are compared and some general principles  are derived. A method for enhancing the radiometric effect and for generating radiometric forces at higher pressures is proposed.

\end{abstract}
  
\section{Introduction}
Crookes' Radiometer, the light mill, is a fragile engine enclosed in a low
pressure vessel and powered by sun light. It was invented by  Victorian scientist Sir William Crookes in 1874 and it was the subject of intense investigation and debate for almost 50 years, being  a laboratory for testing kinetic theory of gases. The radiometer is a small chamber containing a four wing mill mounted on a vertical pivot (see Fig.1). Each wing or vane is typically a 0.1 mm thick , 1 cm$^2$ square mica plate. The vanes are black-lamped on one side and silvered on the other side. When intense light impinges on the vessel,  the mill spins; the black side  becomes hotter than the silvered side because of the larger absorption coefficient of the former. This temperature difference generates a force directed toward the colder surface as  air molecules contained in the vessel strike on the vanes. In fact, air at low pressure exert  forces proportional to the temperature gradient. The light mill can achieve considerable rotational speed when full sun light illuminates the instrument. Before going into the details of the mechanism of action of the radiometer , let us review  few technical data about this instrument. The temperature difference between the hot and the cold side of the vane was measured by Marsh et. al \cite{Marsh} by means of thermocouple and it was found to range from 0.1 to 0.5 K. The total force acting on the four surfaces was measured by Schuster and by O. Reynolds in 1876 \cite{Schuster},  it was measured by suspending the vessel on a pendulum and detecting the torsion of the vessel in the direction opposite to the mill motion, for a rotational speed of 200 revolutions per minute, such force is around $2*10^-6$ Newton, a very small force.  This force decreases when air density inside the vessel is increased. In particular  at atmospheric pressure no motion is observed. Given the simplicity and beauty of the instrument it is of interest to investigate whether by changing one or more of the parameters involved in the problem   radiometric forces could be observed also at atmospheric pressure.
 
Radiometric forces depend on the mean free path of gas molecules, this in turn depends on the density of the gas and on the cross section of the molecules; furthermore they  depend on the value and direction of the temperature gradient. The exact mechanism of action is a complex problem of kinetic theory of gases. A brief description, as it is historically known, is given in the next section. In section III a comparison of two  theories and a derivation of basic principles of radiometric effects are presented. In section IV an improved version of radiometer vane is proposed and an estimate of the enhancement of the force at atmospheric pressure is presented. The results are discussed briefly in the conclusions

\section{Basic mechanism of action}
J.C. Maxwell first pointed out in 1876 that the radiometric force acts only at the edge of the vane being most of the surface inactive, this is because radiometric forces are in substance a heat exchange problem and heat exchange is more efficient at edges, tips and fins.  O. Reynolds \cite{Reynolds} and G. Hettner\cite{Hettner} recognized the existence of a thermal flow of molecules from the cold to the hot side of the vane, the flow velocity is linearly dependent on the temperature gradient $\partial T/\partial x$. The flow  is present near the edge of the vane where a strong gradient exist; in this region air  creeps along the edge  toward the hot surface.  The reaction to this streaming is a force directed opposite to the temperature gradient. Let us  consider a plate  whose section is shown partially in FIG. 2. The plate extends in the $y-z$ (horizontal) plane, the thickness $L$  extends along the $x$ (vertical) direction.  The upper surface  is kept at high temperature $T_h$ and the lower surface is kept at low temperature $T_c$ such that (Th-Tc)= $\Delta T$. The creep force moves  the vane downward; Such force is given below as function of the mass of a molecule of air $m$, its average diameter $\sigma$ the molecule density $n$ and the thickness of the plate $L$:
\begin{equation}
F_{creep}= -\frac{3}{4\sqrt{2}\pi^2}\  \frac{k}{\sigma^2}\ \frac{\Delta T}{L}\ S_v\ ,
\end{equation}
where $k$ is the Boltzmann constant and $S_v$ is the vertical area of the edge.
The creep force is proportional to the surface  of the plate which is parallel to the temperature gradient.  However, being $S_v$ proportional to the thickness  the force is independent of the thickness. When $L<< \lambda$ formula (1) breaks down being the concept of temperature gradient meaningless. In this regime the force is proportional to the thickness of the plate. It is therefore clear that the maximum effective thickness is $L=\lambda$. This is also approximately the thickness chosen by W. Crookes for his radiometer vanes.

There is also another cause of radiometric forces which differs from thermal transpiration. It was  discovered by A. Einstein in 1924 \cite{Einstein} (see also \cite{Loeb}, section 84 where Einstein's original paper is almost entirely translated). Einstein found that radiometric forces act also on surfaces normal to the temperature gradient, i.e. on horizontal surfaces as defined in FIG. 2. 

Einstein's force acts on a tiny portion, a mean free path wide of the vane surface. In FIG.3 a radiometer vane with arm is shown; here the tiny pink strip delimited by the dashed  line is , according to Einstein, the active area of the  vane, the width of the strip is $\lambda$.  Einstein's theory can be summarized as follows : when a large, thin vane is immersed in a region of air where a temperature gradient $\partial T/\partial x$ exists, pressure will be constant in the column of air above the vane, that is the quantity $n(x) T(x)$ is a constant. Away from the plate air must be still, therefore the mass motion of air must be zero and the quantity $n(x)\sqrt{T(x)}$ is a constant. At the edge of the plate there exist  a small transition region of the size of a mean free path in which the pressure is not a constant; in this small region pressure is  higher on the side of the plate facing the higher temperature. A strip of width $\lambda$ on the vane surface will thus suffer a higher pressure from the hot side. The force is proportional to the perimeter of the vane and can be easily calculated by kinetic theory of gases. It is given by:
\begin{equation}
F_{Einstein}= -\frac{1}{2} p  \frac{\lambda^2}{T}\  \frac{\partial T}{\partial x} \ell, 
\end{equation}
where $\ell$ is the perimeter of the vane and $p$ is the gas pressure; this force acts on surfaces normal to the temperature gradient.
Einstein calculates the force  in the case of a vane which is hot on one side and cold on the other as in Crookes radiometer; Einstein inserts into equation (3) the  gradient 
\begin{equation}
\partial T/\partial x = (Th-Tc)/\lambda.
\end{equation}
in reality Crookes vanes have thickness smaller than $\lambda$, however Einstein assumes that $\lambda$ is the minimum effective thickness. The force does not increase for  $L<\lambda$, furthermore the concept of temperature is meaningless below the mean-free-path scale.
As a consequence of this, Einstein finds a radiometric force:
\begin{equation}
F_{Einstein}= -p \lambda \frac{T_h-T_c}{T} \ell,
\end{equation}
    This formula was  tested  in the laboratory by Marsh et al. in  Journal of Optic Society of America (J.O.S.A. \& R.S.I.), 11, 257,1925 and JOSA \& RSI,12, 135, 1926. In these experiments the radiometric force was measured for two kinds of vanes shown in FIG 5 and FIG. 6.  The first is a regular vane similar to Crookes vanes, as shown in FIG. 5; the  second vane has a larger perimeter but equal surface as shown in FIG. 6. The vanes were illuminated on the lamp-blacked side and the force was measured by means of a torsion pendulum.  It was found that on plates with larger perimeter the force was larger. Einstein's formula (5) was verified with good accuracy, the discrepancies being attributed to the difficulty in measuring and maintaining the temperature gradient in the perforated plates. 

\section{Confronting two theories}

Both thermal creep and Einstein's effect contribute to the radiometric forces, however they are commonly viewed as competing theories (see the discussion on section 84, p. 380 of \cite{Loeb}). On closer look  Einstein's theory  implies the existence of a flow of molecules from the cold to the hot side of the gas along the vane edge. In fact in the region at constant pressure, far away from the edge,  the number of impacts on the vane surface is higher at the cold side, thus at the edge of the plate the excess of molecules from the cold side leak along the edge toward the hot side;  therefore thermal creep is included in Einstein's theory. Furthermore Einstein's force is present on the whole surface of a solid body of size $\lambda$ immersed in air; therefore it acts also on surfaces parallel to the gradient. Although a direct calculation of this contribution was never performed, it is expected to be equal to the creep force.
One reason for looking at Einstein' theory with diffidence is that Einstein used a the toy model for his calculation; in this model the gas molecules only move along the three coordinates  with a mean speed $v$. Conversely  the creep force (1) is  calculated by means of non-equilibrium statistics given  by Chapman-Enskog approximation.  It will be shown below that forces acting on surfaces normal to the gradient  can be indeed derived from the Chapman-Enskog approximation. 

Let us consider a system of Cartesian coordinates $xyz$ and a region of air in which a temperature gradient $\partial T/\partial x$ exists. In the presence of the gradient, gas dynamics is  described by the  distribution function:
\begin{equation}
f(v_x, v_y, v_z)\ =\ \left[A + C v_x \left(\frac52 -\beta^2 (v_x^2+v_y^2+v_z^2)\right)\right] e^{-\beta^2(v_x^2+v_y^2+v_z^2)}
\end{equation}
where $A=\left(\frac{m}{2\pi k T}\right)^{3/2}$, $\beta=\sqrt{m/(2kT)}$ and the constant $C$ is given \footnote{Such constant is found imposing uniformity of pressure all through the gas. See \cite{Kennard}} by
\begin{equation}
C=\frac{3 m^2}{4\sqrt{2} \pi^3\sigma^2 n k^2 T^3} \frac{\partial T}{\partial x}\ .
\end{equation}
Let us consider and imaginary surface $S$ normal to the temperature gradient. 
The net momentum flow through the surface is given by 
\begin{eqnarray}
M &=& n m  \int_{-\infty} ^\infty dv_z\ \int_{-\infty} ^\infty dv_y \left( \int_0^{\infty} dv_x v_x v_x f(v_x, v_y, v_z)\right.\nonumber \\ 
  & &  \left.-\ \int_{-\infty}^0 dv_x v_x v_x f(v_x, v_y, v_z)\right)
\end{eqnarray} 
such quantity does not vanish. It is equal to:
\begin{equation}
M = -\frac{3}{\sqrt{2}\pi^2}\ \frac{k}{\sigma^2}\ \frac{\partial T}{\partial x}\ .
\end{equation}

The quantity $M\cdot  S$ is a force acting on a surface normal to the temperature gradient. Formula (8) can be compared to Einstein's radiometric force (2). The latter can be written in the form:
\begin{equation}
F_E\ =\ -\frac{1}{\sqrt{2}\pi}\ \frac{k}{\sigma^2}\ \frac{\partial T}{\partial x}\ (\lambda\ \ell) \ .
\end{equation}
Here the quantity $ (\lambda\ \ell)$ is the active surface as defined above. Einstein's  effect is a force per unit active surface  
\begin{equation}
P_E\ =\ -\frac{1}{\sqrt{2}\pi}\ \frac{k}{\sigma^2}\ \frac{\partial T}{\partial x}\ \ .
\end{equation}
It can be seen that eq.(8) and eq.(10) are similar, their ratio being  just  $3/\pi$. Recalling that Einstein's formula is  an approximation  the agreement of the two theories is excellent.

Whatever the approach used,   at the edge of the radiometer vane the force acts on surfaces both normal and parallel to the gradient. The active portions of surface are strips of material a mean free path wide  In particular a small solid body of size $\lambda$  experiences a radiometric force all over its surface. In the case of a small parallelepiped the force on horizontal surfaces can be calculated with good approximation by equation (10), while the force on vertical  surfaces is calculated by eq.(1). Roughly The total radiometric force is given by the sum of these contributions.

As a general principle,  radiometric forces arise in the presence of: 1. a temperature gradient, 2. a solid body. In the absence of solid bodies no force acts and the gas is stationary. The shape of the solid body is not relevant as long as the size of the body is smaller or equal to the mean free path. For bodies larger than $\lambda$ edges are relevant; here only portions of surface a mean free path wide are active. For bodies smaller than $\lambda$ the force decrease with the surface. The radiometric force is always directed opposite to the temperature gradient and therefore opposite to the  heat flow.

\subsection{Improved radiometer vane}

From formula (9) and from the experiments  of Marsh et al. it is clear that the radiometric force can be enhanced by increasing the perimeter. 
The force should be  greatly enhanced when the vane is perforated with a large number of small apertures, like for instance  in Fig 4. Here the upper view of the radiometer vane shows hexagonally packed arrays of apertures. The width of the strips of material in between the apertures is  $\lambda$ on average. 
Apertures  of size  lambda or less, spaced so as to open approximately 50\% of the entire surface should maximize the force. Since Einstein's effect is caused by the presence of a transition region between closed and opened spaces, an equal distribution of open and closed spaces in Fig. 4 should generate a transition region extending over the whole area of the vane.  In this case the whole of the closed surface would be radiometrically active. Einstein's force can be easily calculated for this vane configuration.
Equation (4) can be rewritten in the form:
\begin{equation}
F_{Einstein}= -n k (T_h-T_c) \ \left(\lambda \ell \right)
\end{equation}
since $p=n k T$. The quantity $\lambda \ell$ inside the parenthesis in Equation (11) is what we  defined above as the closed  surface. Equation (11) can be rewritten in the compact form  
\begin{equation}
F_E= -n k (T_h-T_c) S_h\ .
\end{equation} 
In the case of a closed surface equal to 50$\%$ of the total area $A$ of the vane (closed + open surface), one finds a force per unit area of the vane
\begin{equation}
P_E= -\frac 12 n k (T_h-T_c)
\end{equation}
Formulas (11), (12) and (13) are valid when the temperature gradient is given by Equation (4), thus the thickness $L$ of the vane should be $\lambda$. Were the thickness to be larger than $\lambda$ the temperature gradient would be smaller and the force would be smaller. As explained in the first section of this paper $\lambda$ is the minimum effective thickness.
Formula (13) accounts for the force acting on the perforated surface normal to the gradient. However  also the surfaces parallel to the gradient are  active.  The force on this surface can be calculates by Equation (1) for thermal creep. 
When a temperature differential is applied to the a perforated plate, like the one in Fig. 4,  air will stream through the  apertures which are in fact channels with a finite length. The streaming is directed from the cold to the hot surface. If the channels are very long, the flow will be governed by Poisseuille law for gas flows through long tubes.  Therefore, in order to maximize the creep force the channels should be short. The minimum effective channel length is $\lambda$. Below this value the force decreases with the channel length.  Therefore the thermal creep force is maximized when the vane thickness is $\lambda$ and the number of channels is as large as possible. 

The vane thickness, is the crucial factor for maximizing  radiometric forces due to both Einstein's effect and thermal creep. While the size and shape of the apertures can vary, the thickness of the plate must be as close as possible to $\lambda$.

Let us calculate the creep force acting on a perforated plate of this thickness Equation (1) can be rewritten as:
\begin{equation}
F_v= -\frac{3}{4 \pi} n k (T_h - T_c) S_v
\end{equation}
The vertical surface $S_v$, is a function to the thickness   the number of apertures $j$, and the perimeter $f$ of a single aperture, according to the following formula:
\begin{equation}
S_v= 4 b L + j f L
\end{equation}
Even if the plate is only a mean free path thick, $S_v$ can be large, assumed the number of apertures is large.
In the case of circular apertures of radius $R$ and with a 50\% open surface one finds a force per unit area of the plate:
\begin{equation}
P_v= -\frac{3}{4 \pi} n k (T_h - T_c) \frac{\lambda}{R}
\end{equation}
 
According to the general principles governing radiometric forces, a small body of size $\lambda$ (or smaller) suffers a radiometric force on the whole of its surface. The thin perforated  plate in Fig. 4 is  a sort of aggregate of small solid bodies  with length scale equal to $\lambda$.

Using the vane configuration just described, radiometric forces can be in principle  observed at higher gas densities.
With the shrinking of the mean free path, all features of the vane must shrink. In particular the thickness of the vane and  the size and spacing of the apertures must be smaller. At  atmospheric pressure the mean free path is about 70 nm, such length scale  poses several engineering challenges. On the other hand, nanotechnology allows for fabrication of structure as small as 20nm.  Nanometer scale perforation would greatly increase the active surface. For instance, at atmospheric pressure a square vane of side b=1m and thickness L=70 nm perforated with circular holes of 70 nm diameter and with an open area of 50\% would have a total active area of $2.5 $m$^2$. Without perforation the same vane would have an active area of about $6*10^{-7}$m$^2$. Thus, by means of perforation, the active area is increased by a factor $10^7$. Being the force linearly dependent on the active area, it is expected to increase by the same factor. It should be noted that by opening a large number of channels through the vane the heat transport from the hot to the cold regions would be enhanced as well. Therefore  it is expected that a larger amount of radiant energy per unit time is necessary to maintaint the temperature gradient in a perforated plate.

\subsection{Conclusion}
Radiometric forces are an interesting thermal effect explained by gas  kinetic theory.  Such forces act on small bodies immersed in a gas where a temperature gradient is present. The forces  are linearly dependent on the temperature gradient and  are directed as the heat flow. Although the forces observed in Crookes radiometer are small, it is possible to enhance them by means of perforation of the vane. Such vane would behave as an aggregate of small bodies. When the size of the apertures and the thickness of the vane is sufficiently small the forces can be significant even at high gas densities. In particular nanotechnology could permit for an enhancement of the force by a factor $10^7$ with respect to the typical order of magnitude. Such force could be observable at atmospheric pressure.

\vfill\eject

\begin{figure}[ht]\unitlength1cm
\begin{picture}(10,10)
\put(5,0){\epsfig{file=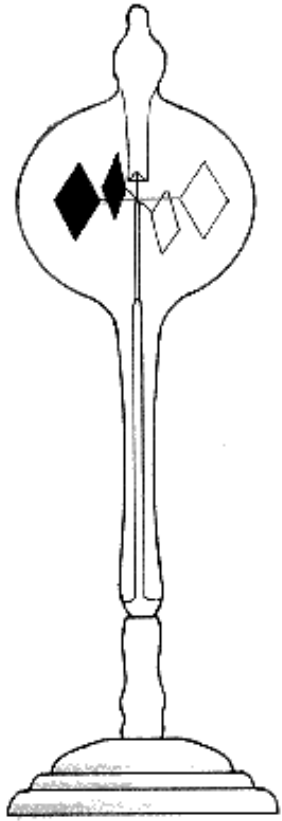,width=4cm}}
\end{picture}
\caption{Crookes radiometer}
\end{figure}

\begin{figure}[ht]\unitlength1cm
\begin{picture}(10,10)
\put(0,5){\epsfig{file=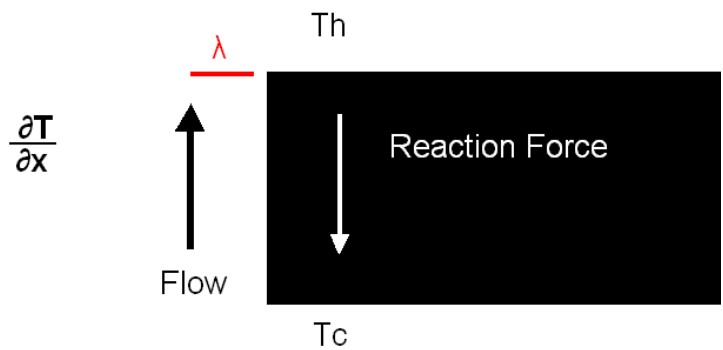,width=10cm}}
\end{picture}
\caption{Thermal creep along the edge of the radiometer vane, the flow is parallel to the temperature gradient and it is directed opposite to the heat flow. The reaction to the creep is a radiometric force directed as the heat flow}
\end{figure}

\begin{figure}[ht]\unitlength1cm
\begin{picture}(10,10)
\put(-8,2){\epsfig{file=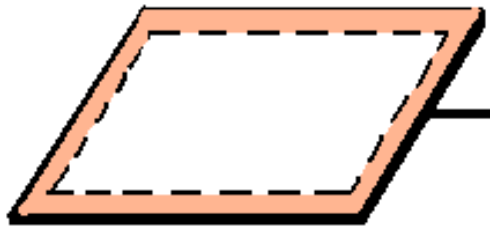,width=20cm}}
\end{picture}
\caption{The active region of the vane surface according to Einstein's theory}
\end{figure}

\begin{figure}[ht]\unitlength1cm
\begin{picture}(10,10)
\put(0,0){\epsfig{file=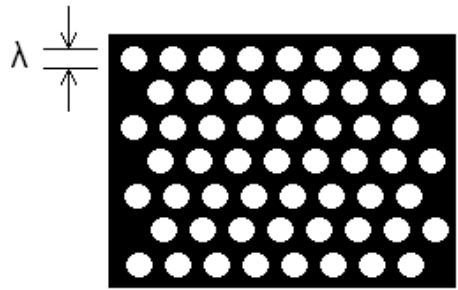,width=15cm}}
\end{picture}
\caption{An improved radiometer vane. The vane has hexagonally packed apertures. The open area is 50\% of the whole vane surface, the average distance between the holes is equal to the mean free path $\lambda$. The thickness of the vane (not shown in the figure) is exactly  $\lambda$. Such geometry maximizes Eintein's effect and thermal creep}
\end{figure}

\end{document}